# Toward the Identification of Atomic Defects in Hexagonal Boron Nitride: X-Ray Photoelectron Spectroscopy and First-Principles Calculations

Gabriel I. López-Morales[1,2,3], Nicholas V. Proscia[1,4], Gustavo E. López[2,3], Carlos A. Meriles[1,4], Vinod M. Menon[1,4]

[1]*Department of Physics, The City College of the City University of New York, New York, NY 10031 USA*
[2]*Department of Chemistry, Lehman College of the City University of New York, Bronx, NY 10468, USA*
[3]*Ph. D. Program in Chemistry, The Graduate Center of the City University of New York, New York, NY 10016, USA*
[4]*Ph. D. Program in Physics, The Graduate Center of the City University of New York, New York, NY 10016, USA*

**Abstract:** Defects in hexagonal boron nitride (hBN) exhibit single-photon emission (SPE) and are thus attracting broad interest as platforms for quantum information and spintronic applications. However, the atomic structure and the specific impact of the local environment on the defect's physical properties remain elusive. Here we articulate X-ray photoelectron spectroscopy (XPS) and first-principles calculations to discern the experimentally-observed point defects responsible for the quantum emission observed in hBN. XPS measurements show a broad band, which was deconvolved and then assigned to $N_BV_N$, $V_N$, $C_B$, $C_BV_N$, and $O_{2B}V_N$ defect structures using Density Functional Theory (DFT) core-level binding energy (BE) calculations.

Hexagonal boron nitride (hBN) is a common wide band-gap semiconductor considered for the development of quantum communication devices[1, 2] because of a) its van der Waals nature, which allows mechanical exfoliation of the bulk crystal into atomically-thin structures, b) its ability to host defects that exhibit bright single-photon emission (SPE) at room-temperature,[3] and c) its wide band-gap (~6 eV), which makes it possible to probe defects without generating excitonic emission.[4] Additionally, hBN defects have shown interesting optical and magnetic properties, such as strain-activated and tunable SPE, and spin-dependent intersystem crossing (ISC).[5–7]

Despite extensive investigation of these defects, their specific characteristics and atomic structures are not well understood.[8] Theoretically, the fact that most of the proposed defects have similar properties makes it hard to unambiguously assign point defects to experimentally observed properties.[9–11] Another theoretical complication arises from a limitation of Density Functional Theory (DFT), which cannot reliably predict the band-gap nor defect energy levels in wide-gap semiconductors.[10, 11] Calculations on electronic transitions and optical responses must therefore be taken at face value. Defects in hBN are also difficult to probe experimentally due to their dynamic behavior, the limited reproducibility between observed results[5, 7, 12, 13] and the variations between sample preparation or treatments.[1, 14, 15]

Techniques for elemental and compositional analysis include XPS, time-of-flight secondary ion mass spectrometry (ToF-SIMS) and, electron paramagnetic resonance (EPR). While EPR has been used to identify defects in semiconductors,[16, 17] it is limited to the study of defects with magnetic properties and available in relatively large numbers. On the other hand, ToF-SIMS provides enhanced atomic detection limits,[18] but is destructive to the sample surface. To experimentally construct a complete picture of the impurities that are experimentally observed in hBN using a non-destructive and accurate technique, XPS measurements were performed. Assignments of the experimentally observed peaks were achieved using first-principles calculations.

XPS measurements were carried out on two samples, hBN mechanically-exfoliated from the bulk crystal, and hBN grown by chemical vapor deposition (CVD) on copper foil; both transferred onto $SiO_2$ substrates (see **Supporting Material**). The binding energies (BEs) obtained for boron 1s (B1s), corresponding to the exfoliated and the CVD-hBN, are shown in **Figure 1(a).** The B1s peak for the CVD-hBN is broadened and shifted with respect to that corresponding to the exfoliated hBN. These changes suggest new B-related bonding in the CVD-hBN, since core-level BEs are known to change as a function of the chemical environment surrounding each atom.[19]

Both samples were also inspected by means of photoluminescence (PL) measurements, which were performed via a custom-built confocal setup (excitation source of 460 nm), with an 80/20 splitter that allows for real-time spectral analysis and single-photon counting. **Figure 1(b), (c)** show the PL spectroscopy of these samples. The exfoliated hBN did not show presence of bright emitters, whereas the CVD-sample shows a much brighter background and photo-stable emission at specific locations. The opto-electronic properties of hBN between the samples are thus inherently different, and so are their bonding profiles. These results suggest that part of the new bonding present in the CVD-sample is responsible for the observed differences in emission.

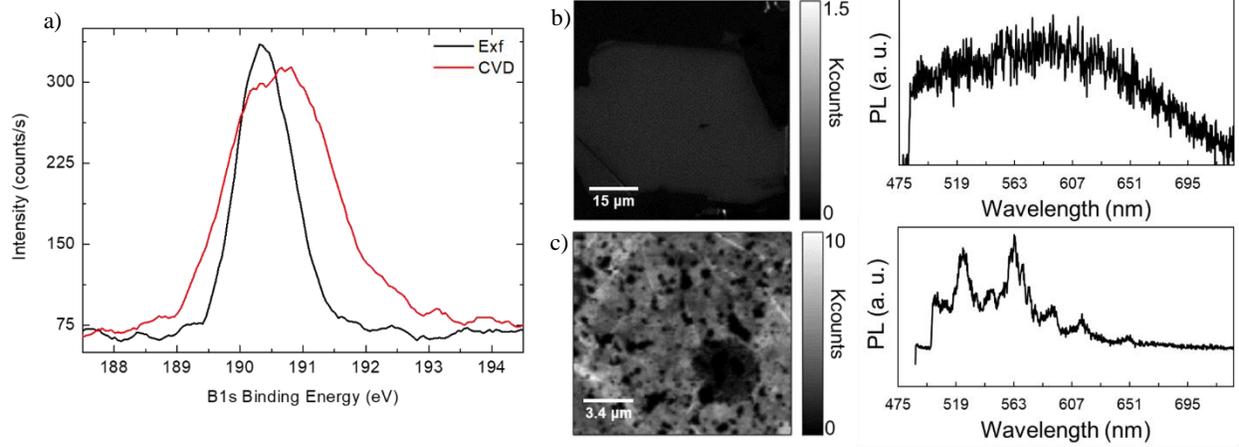

**Figure 1**. (Color online) XPS and PL spectroscopy of exfoliated and CVD-grown hBN. The B1s electron BE peak corresponding to the CVD-hBN shows increased broadening with respect to that of the exfoliated sample, suggesting new B-bonding in CVD-hBN. As seen from the confocal images of (b) an exfoliated hBN flake, and (c) CVD-grown hBN, and their respective optical spectra (right), the samples have markedly different optical properties.

The atomic species of interest (B, N, O and C) were further analyzed to probe the different bonding present in both the CVD- and exfoliated hBN. Deconvolutions were performed based on best fits seeking consistency between peaks and noting that each deconvolution element must correspond to a specific type of bonding within the sample matrix.

To further understand the observed XPS spectra from hBN samples, XPS measurements on specific defects were simulated using DFT in the generalized gradient approximation (GGA). As outlined elsewhere[20–25], there are different theoretical approaches that can be used to calculate core-level BEs using DFT, *e.g.*, the self-consistent field (ΔSCF) method, the complete screening picture method, and the Slater-Janak transition state (SJTS) method (**Supp. Mat.**).

In calculating core-level BEs within the SJTS method, the Kohn-Sham core eigenstates are recalculated for optimized defect structures via the initial- and final-state approximations, with the excited core electron added to the conduction band minimum (CBM), creating an effective hole in the core of the corresponding atom. Given its convenience in terms of accuracy and computational cost, this method was implemented using the Vienna *ab initio* Simulation Package (VASP) within the projector-augmented wave (PAW) method. Perdew-Burke-Ernzerhof (PBE) functionals were used to account for the exchange-correlation interactions. Additionally, core-level BEs of the different species in each sample and their corresponding shifts[20–25] were obtained using Equation 1:

$$BE = \frac{[(E^i_F - \epsilon^i_C) + (E^f_F - \epsilon^f_C)]}{2} \tag{1}$$

where $E^k_F$ is the Fermi level of state $k = i, f$ and $\epsilon^k_C$ represents the core-eigenstates. This expression includes corrections for the Fermi level and the local potential at the vacuum region.

**Figure 2** shows the calculated BEs for the B1s electrons in pristine hBN, and the N1s electrons in the $O_B$ defect (shown as an example) versus supercell size. The BEs did not show a large fractional change, nor any specific trend, as a function of the supercell size in either case. This means that using a smaller cell size is as valid as using a larger one and thus, a 4x4 supercell with a vacuum region of 15 Å was chosen for all calculations. The obtained experimental data was used as a reference[26], and theoretical BEs were used to assign deconvoluted XPS peaks to specific defect structures in hBN.[27]

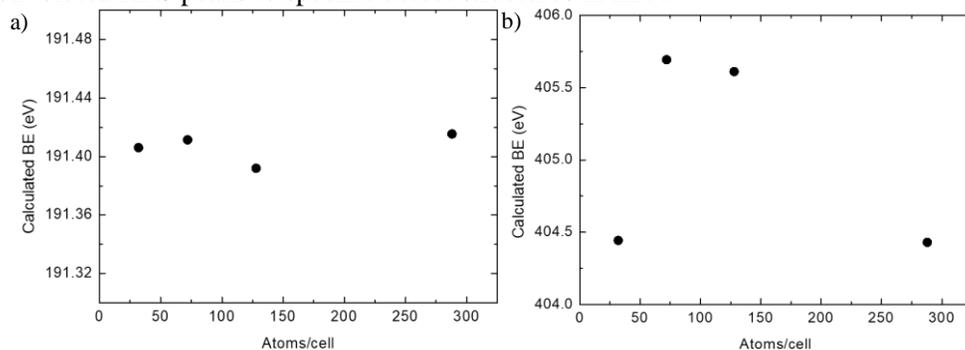

**Figure 2**. Calculated BEs as a function of cell size for two different structures. a) B1s binding energies in pristine hBN, b) N1s binding energies in the $O_B$ defect structure. The calculated BEs do not follow a trend as cell size increases.

Deconvolution of the B1s and N1s BE peaks are shown in **Figure 3**. In the case of CVD-hBN, the fits show peaks not present in the exfoliated hBN, suggesting different B- and N-related defect profiles between the samples.

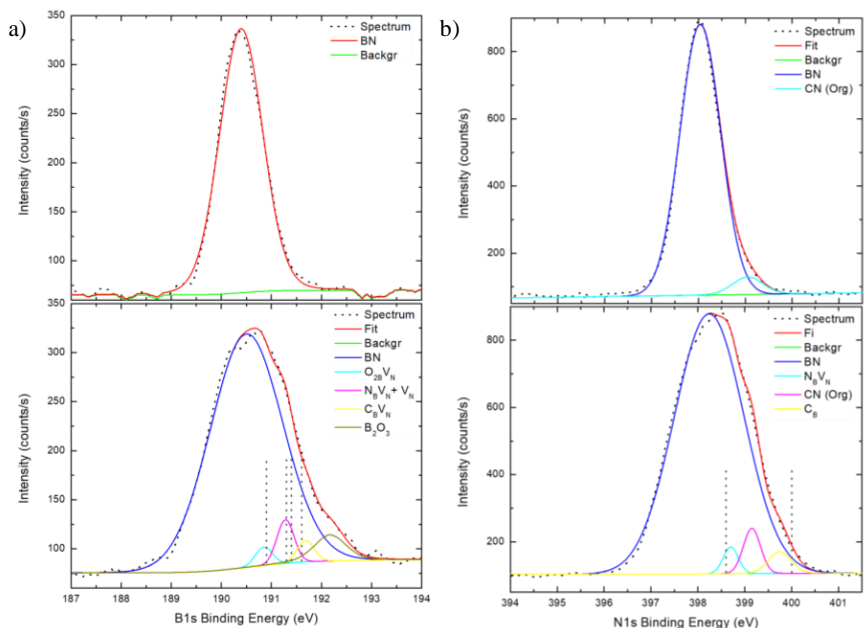

**Figure 3**. (Color online) Deconvoluted XPS (a) B1s and (b) N1s peaks from exfoliated hBN (top) and CVD-hBN (bottom). Black dotted lines represent experimental XPS spectra, while red lines correspond to the fits obtained from the sum of Gaussian-Lorentzian curves centered at different energies. DFT-calculated BEs for each defect are shown as dashed vertical lines.

DFT calculations on several defect structures (details found in the **Supp. Mat.**), show that the $N_BV_N$ (**Figure. 5(b)**) corresponds to a B1s BE of 191.4 eV, whereas for the $V_N$ (**Figure 5(a)**) the B1s BE is 191.5 eV. Among the studied defects, these were the only intrinsic ones found to fit within the obtained XPS peaks. Extrinsic defects such as the $C_B$ (**Figure 5(d)**), with N1s BE of 400.1 eV, the $C_BV_N$ (**Figure 5(c)**), and the $O_{2B}V_N$ (**Figure 5(e)**), with B1s BEs of 191.7 eV and 191.0 eV, respectively, were also found to fit consistently inside the obtained BE spectra. The maximum standard deviation in supercell size assessments was found to be 0.5 eV and was taken as the acceptance criteria for assigning defects to specific peaks. This is in consideration of the errors from cell local potentials, defect-defect interactions, surface effects and the reference energy range for each XPS peak.

Based on the outlined results, the peak at 191.3 eV in the B1s BE peak for the CVD-grown hBN can be attributed to either $V_N$ or $N_BV_N$ defects. However, due to resolution limitations and the weak signal amplitude observed for both the N1s and B1s lines, this peak cannot be further resolved into the $V_N$ and $N_B V_N$ contributions. A BE peak that could correspond to the $N_B V_N$ was also observed in the N1s peak (398.7 eV), while the $V_N$ defect could not be corroborated in this fashion (**Supp. Mat.**). Given the proximity between calculated BEs for $V_N$ and $N_B V_N$, both defect contributions are included inside the peak centered at 191.3 eV.

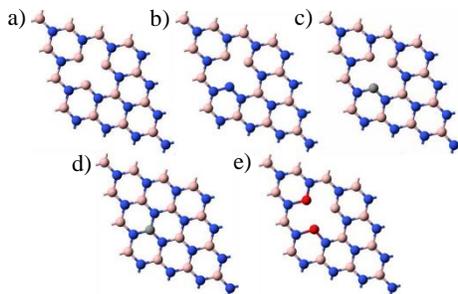

**Figure 4**. (Color online) Defects structures identified in CVD-grown hBN. Based on XPS measurements and DFT calculations, the a) $V_N$, b) $N_B V_N$, c) $C_B V_N$, d) $C_B$, and e) $O_{2B} V_N$ defects have been identified in the CVD-hBN. On the other hand, $N_V$ and $N_B V_N$ defects are likely to be present in the exfoliated hBN, with concentrations lower than the detection limit for XPS. Boron, nitrogen, carbon and oxygen atoms are presented by pink, blue, grey, and red spheres, respectively.

Significant C- and O-bonding in hBN was not observed for the exfoliated sample. Interestingly, due to new bonding also observed in the C1s and O1s BE peaks for the CVD-hBN with respect to the exfoliated sample, some of the new peaks arising from the broad shoulder centered at ~191.8 eV in the B1s BE peak for the CVD-hBN could be attributed to C- and O-related defects. Based on DFT results for the C1s, N1s and O1s peaks (**Supp. Mat.**), peaks centered at 190.9 eV, and 191.6 eV in the B1s BE could correspond to the $O_{2B} V_N$, and $C_B V_N$ defects, respectively. Additionally, a peak centered at 399.8 eV in the N1s BE peak, and one centered at 286.4 eV in the C1s BE peak, suggest the presence of $C_B$ centers in the CVD-grown sample. This defect however, provides only one defect orbital inside the band-gap and is thus unlikely to result in sharp PL.[9,10] Finally, the peak centered at 192.1 eV in the B1s peak is assigned to boron oxide present on the surface of the hBN layer, while other peaks identified as having organic nature are likely due to impurities from the transfer process.

The obtained results suggest that $V_N$, $N_B V_N$, $C_B V_N$, and $O_{2B} V_N$ defects are responsible for most of the observed emission in CVD-hBN. However, $N_B V_N$ and $V_N$ are likely to play an important role in both samples, since bright emission has been observed in exfoliated hBN after argon treatment,[8] which in principle should not incorporate extrinsic impurities. It should be added that recent calculations on the electronic properties of some of these defects have shown a) zero-phonon lines (ZPLs)[9,10] of 2.08, 1.90, and 2.12 eV, b) Huang-Rhys (HR) parameters[10] of 1.66, 6.74 and 4.49, and c) symmetry corresponding to $C_{2v}$, $C_s$, and $C_{2v}$, for the $C_B V_N$, $O_{2B} V_N$, and $N_B V_N$, respectively. Some of these details have also been experimentally corroborated, and the results presented herein support reported observations. Lastly, the calculated defect concentrations based on XPS data (on the order of 0.8-2%) suggest complex mechanisms of defect activation, likely due to: mixing of defect-related intra-band states (defect-defect interactions), charge-carrier dynamics, and relative energy shifts due to local strain. On the other hand, such high concentration of defects could also be a result of averaging areas of highly-defected hBN over the whole sampling area. It is noted that these observations could provide understanding for the bright emission present along the edges and grain boundaries of hBN samples, where defect concentration is highest. Additionally, it could also serve as a possible explanation for the bright background present in the CVD-grown hBN as compared to the exfoliated sample.

In conclusion, atomic characterization of hBN samples grown under different conditions has been presented through the scope of confocal spectroscopy and XPS, thus helping to create a more complete picture of the defects of interest, their optical properties, and the effect of growth conditions on their profiles. The samples showed inherently different emission, and XPS results confirmed that their chemical environments were indeed distinctive. Experimental XPS peaks obtained from both samples were fit by deconvolving the electron BE spectra. Deconvolutions were assigned to $N_B V_N$, $V_N$, $C_B$, $C_B V_N$, and $O_{2B} V_N$ defect structures, based on the chemical bonding within the atomic species and DFT core-level BE calculations. As a computationally-inexpensive approach, the method described herein provides insight into the theoretical BE calculations and the assignment of defect structures in 2-dimensional (2D) semiconducting materials.

C.A.M acknowledge support from the National Science Foundation (USA) through grants NSF-1619896 and NSF-1401632, and from Research Corporation through a FRED award. G.I.L.M. and N.V.P. acknowledge support from CREST IDEALS, NSF-1547830.